# Global Inverse Design Across Multiple Photonic Structure Classes Using Generative Deep Learning

*Christopher Yeung[1,2], Ryan Tsai[1,3], Benjamin Pham[1], Brian King[1], Yusaku Kawagoe[1], David Ho[1], Julia Liang[1], Mark W. Knight[2], and Aaswath P. Raman[1,*]*

[1]Department of Materials Science and Engineering, University of California, Los Angeles, CA 90095, USA
[2]Northrop Grumman Corporation, Redondo Beach, CA 90278, USA
[3]Harvard-Westlake School, Los Angeles, CA 90077, USA
*Corresponding Author: aaswath@ucla.edu

**Abstract:** Understanding how nano- or micro-scale structures and material properties can be optimally configured to attain specific functionalities remains a fundamental challenge. Photonic metasurfaces, for instance, can be spectrally tuned through material choice and structural geometry to achieve unique optical responses. However, existing numerical design methods require prior identification of specific material-structure combinations, or device classes, as the starting point for optimization. As such, a unified solution that simultaneously optimizes across materials and geometries has yet to be realized. To overcome these challenges, we present a global deep learning-based inverse design framework, where a conditional deep convolutional generative adversarial network is trained on colored images encoded with a range of material and structural parameters, including refractive index, plasma frequency, and geometric design. We demonstrate that, in response to target absorption spectra, the network can identify an effective metasurface in terms of its class, materials properties, and overall shape. Furthermore, the model can arrive at multiple design variants with distinct materials and structures that present nearly identical absorption spectra. Our proposed framework is thus an important step towards global photonics and materials design strategies that can identify combinations of device categories, material properties, and geometric parameters which algorithmically deliver a sought functionality.

A central challenge in contemporary materials and photonics research is understanding how intrinsic materials properties can be optimally combined with nano- or micro-scale structuring to deliver a target functionality. Metasurfaces, for instance, hold the potential to become a vital component for many next-generation optical technologies due to their ability to manipulate the propagation of light within an ultracompact footprint.[1] More broadly, by leveraging subwavelength nanostructures and the intrinsic dispersion of constituent materials, tailored changes in the amplitude and phase of incident wavefronts can be precisely engineered, along with desired spectral characteristics. This new level of control has enabled and accelerated critical developments in fields such as flat optics,[1-3] quantum communications,[4] and holography.[5,6] However, our ability to meet increasing demands in the performance of metasurfaces, and photonic structures in general, faces roadblocks due to the complexity of the materials and structural design spaces that are currently accessible.

From the perspective of a researcher or practitioner in the field, enabling a desired set of optical characteristics today typically involves a prior understanding of the capabilities of different categories of devices or nanostructures. For instance, ultra-strong field confinement may lead one to start with a plasmonic architecture, while high transmission applications would lead one to ensure the use of materials that present low extinction coefficients in the wavelength range of operation. Designing photonic structures that meet application-specific objectives thus entails identifying the ideal intersection of material properties, structural composition, and fabrication process (or device class), as specific combinations are more likely to yield desired functional characteristics. It is only once a device or photonic structure category has been identified that numerical optimization methods typically enter the picture to optimize and refine performance characteristics.

Conventional optimization methods, which rely on numerical simulations that solve Maxwell's equations, have shown remarkable capabilities in designing nanophotonic structures and are now commonly used.[7] However, they can be computationally costly and are often intractable for large-scale designs or high-dimensional design spaces.[8,9] As a result, data-driven approaches based on machine learning (ML) have been extensively explored in order to tackle challenging photonics design problems.[10,11] Current state-of-the-art machine learning methods involve training neural networks to learn the underlying relationships between photonic structures and corresponding optical phenomena. A trained neural network can, in principle, instantaneously generate designs with substantially lower computational costs than optimization-based methods. A wide range of neural network and machine learning architectures have been investigated for the design and characterization of materials.[12-15] In the photonics context, one-dimensional (1D) tandem networks were used to design core-shell nanoparticles,[16] multilayer thin films,[17] and supercell-class metasurfaces.[18] However, such network architectures are only applicable to simple photonic structures for which geometric and material properties can be described by a vector of discrete parameters.[19] In contrast, photonic devices with complex freeform geometries cannot be well-represented by discrete variables, but offer the potential to achieve new functionalities and greater device performance.[20] For these structures, image-based generative networks have successfully designed various types of metasurfaces, including ones with silver,[21] gold,[22] or silicon[23] meta-atoms and other topological features. Further studies have combined image-based ML with optimization algorithms to yield even greater model performance.[24,25]

Despite the significant progress in image-based photonics design, existing studies are limited to designing the two-dimensional structural topology (or geometry) for a single class of metasurface or nanophotonic structure. In addition, the material properties and out-of-plane parameters (*e.g.*, layer thicknesses) of the explored structures are typically held constant. The central limitation identified earlier remains: prior knowledge of which category of structures or devices may deliver a specific functionality is needed before initiating the optimization procedure (whether machine learning-based or otherwise). However, human intuition on the optimal nanostructure category — the initial conditions for a numerical optimization procedure — can often go awry when faced with competing design goals. Thus, a unified 'global' materials and photonics inverse design approach that can define both the materials and structure (beyond 2D) across multiple classes of photonic structures has yet to be demonstrated, but could fundamentally change how we approach the design and optimization of photonic structures and metamaterials. Moreover, such a capability could prove critical to the design of nonlinear and phase-changing platforms where optical response depends heavily on material composition and fabrication process.[26]

In this study, we present an image-based deep learning framework for the inverse design of photonic structures across multiple materials and device categories. Our approach combines the advantages of material property and structural parameter prediction enabled by 1D tandem networks, with the freeform design capabilities of image-based deep learning. This is accomplished through a versatile image-encoding technique where material and structural parameters such as refractive indices, plasma frequencies, layer thicknesses, resonator geometries, and metasurface classes are embedded within the discrete 'RGB' channels of colored images. Although we show multiparametric encoding through different shades of color in a 3D array (as an initial demonstration), we note that this information can also be encoded via higher-dimensional matrices or data structures that extend beyond the 'RGB' color system. The encoded images are used to train a customized conditional deep convolutional generative adversarial network (cDCGAN), which we evaluate by inputting a variety of target absorption spectra. In response to the input spectra, the network generates corresponding metasurface designs that are validated through full-wave electromagnetic (EM) simulations. To determine network accuracy, performance, and generalizability, the simulated spectra are compared to the input targets. Through this process, we demonstrate that the network simultaneously optimizes the material properties and 2.5D structuring across multiple classes of metasurfaces, thus validating the feasibility of a global inverse design framework that accounts for all the parameters which govern the optical behavior of photonic structures. We note that 'global' in this context refers to the network's ability to perform a global search within the surveyed design space,[8,26] which includes material properties and freeform topology, but the network does not guarantee that the final generated device is globally optimal.

## Results and Discussion

We consider two classes of absorbing metasurfaces in developing and demonstrating our inverse design approach (Figure 1a). First, we consider metal-insulator-metal (MIM) structures, where a thin dielectric layer is sandwiched between two metal layers (one uniformly deposited and the other lithographically patterned). This class of metasurface exhibits a relatively broad Lorentzian-shaped absorption response supported by each individual resonator, which renders this type of structure highly-amenable to thermal emission and energy harvesting applications.[27,28] Next, we consider hybrid dielectric metasurfaces with a metal film substrate, which take advantage of a cavity effect to produce an asymmetric, narrow-band Fano resonance that is well-suited for optical sensing and detection.[29]

As seen in Figure 1b, the first step of our encoding method involves capturing the planar geometries ($G$) and material properties of the metasurface resonator ($M$), followed by the thicknesses of the dielectric layer (*T*), for both MIM and hybrid dielectric metasurfaces. We then encode *G*, *M*, and *T* into the red, green, and blue channels of a colored image. Within our encoding scheme, the red-channel represents the plasma frequency ($M=\omega_P$) and shape of the metal resonator in an MIM structure. The green-channel represents the real refractive index ($M=n$) and shape of the dielectric resonator in a hybrid dielectric structure. The remaining pixels in the blue-channel are used to define the thickness of the dielectric layer (in nanometers) for both metasurface classes. Thus, a red-blue color scheme indicates MIM structures while green-blue indicates hybrid dielectric structures (red-green image combinations are undefined). With this strategy, in addition to representing resonator geometry, different colors on an image can be used to describe unique combinations of material and structural parameters, which in turn yield significantly more variation in achievable optical responses than single-material approaches.

Though the described material properties can be denoted by individual values instead of entire image channels, the presented channel-encoding method offers several key advantages. First, it combats the well-known noise-related artifacts found in image-based ML techniques such as generative adversarial networks (GANs)[8,21] by ensuring that the encoded properties are appropriately weighted towards the network's final predictions. A detailed analysis of models trained on several property-encoded neurons versus models trained on whole image channels is found in the Supporting Information. Additionally, in principle, our approach only requires small modifications to the input dimensions of an existing model (*e.g.*, changing from a 64×64 to 64×64×3 matrix), which allows us to leverage existing model optimization and training techniques without significantly increasing training costs. Furthermore, the presented method is capable of representing spatially-varying material properties along the entire physical structure, which enables the design of 3D or complex gradient-index and metal alloy-based structures that are, in principle, amenable to existing fabrication methods.[49] A demonstration of this design capability is shown in Figure S6.

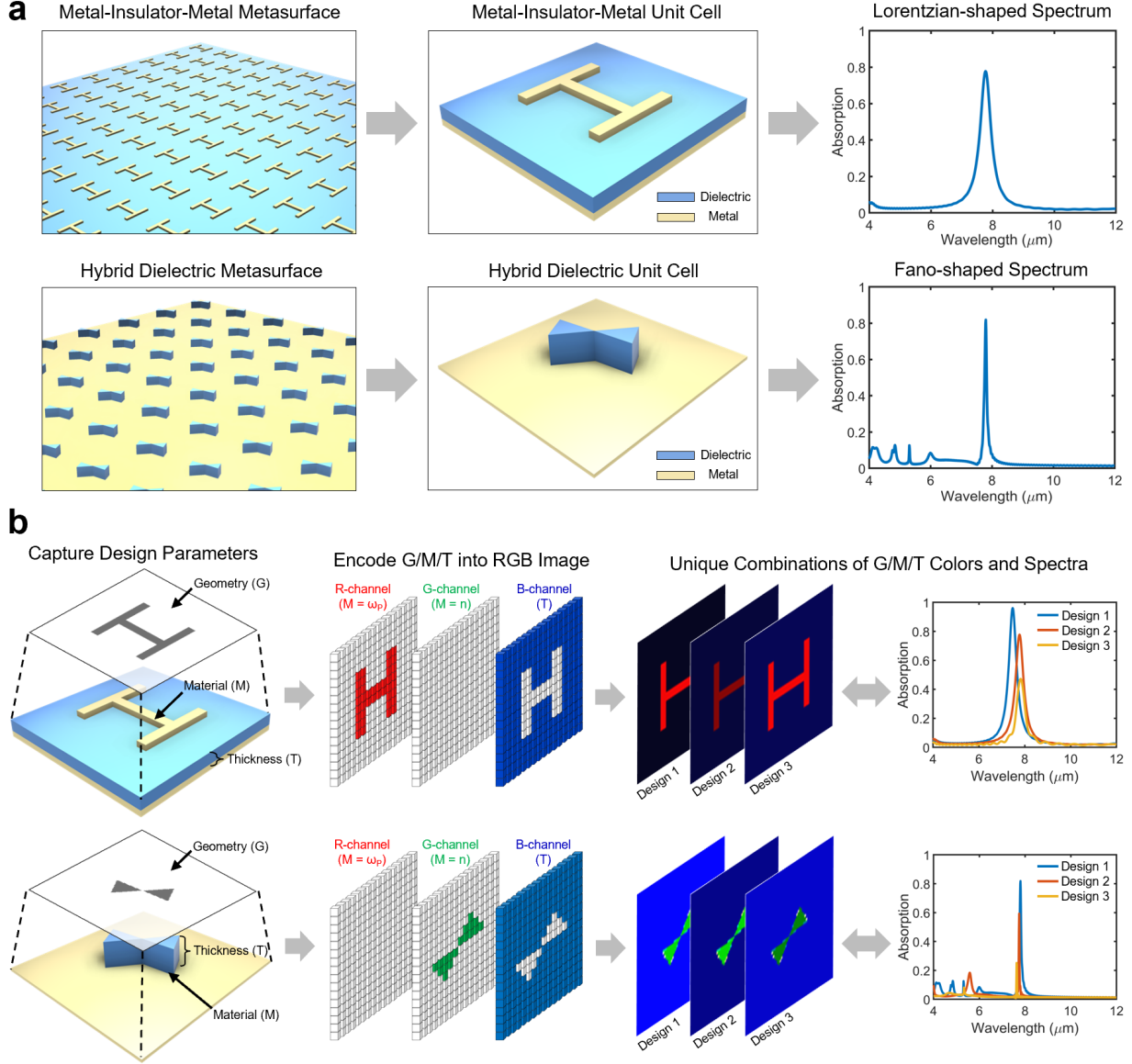

**Figure 1.** (a) MIM and hybrid dielectric metasurfaces with Lorentzian-shaped and Fano-shaped absorption responses, respectively. (b) Representing distinct classes of metasurfaces as color-encoded images. Metasurfaces are converted into images representing their planar geometries. Material properties, thickness values, and metasurface class are encoded into the images as various shades of color, allowing more degrees of freedom for metasurface design.

Our training dataset consists of 20,000 metasurface unit cell designs, represented as image-vector pairs, derived from seven shape templates: cross, square, ellipse, bow-tie, H, V, and tripole-shaped. Detailed information regarding these designs are found in Figure S1 of the Supporting Information. MIM and hybrid dielectric structures are captured within 3.2×3.2 $\mu m^2$ and 7.5×7.5 $\mu m^2$ unit cells, respectively. Each design was converted into a 64×64×3 pixel 'RGB' image using the rules established above. A single pixel therefore corresponds to a minimum feature size of 50 nm (MIM) and 120 nm (hybrid dielectric), which is well-within feasible fabrication range.[21,50] Furthermore, we employed a Gaussian filtering post-processing procedure (described in the Supporting Information) to enhance device performance and fabricability. Finite-difference time-domain (FDTD) simulations were performed on the designs (Lumerical) to obtain an 800-point absorption spectrum vector (from 4-12 $\mu$m) for each structure. Low quality designs (defined in the Supporting Information) were removed from the training set to maximize the model's utility and performance.[30] Figure S2 illustrates the peak absorptions and resonance wavelengths of the spectra represented in the final training dataset.

During the color-encoding step, the Drude model plasma frequencies of the metal resonators ($\omega_P$=1.91 PHz for gold[31], $\omega_P$=2.32 PHz for silver[32], and $\omega_P$=3.57 PHz for aluminum[33]) were used to encode the red channel, and the real refractive indices of the dielectric resonators ($n$=2.41 for zinc selenide[34], $n$=3.42 for silicon[35], and $n$=4.01 for germanium[35]) were used to encode the green channel. The encoded material properties are based on optical constants from the same mid-infrared wavelength range as the simulations. A range of dielectric thickness values (100 nm to 950 nm) were used for the blue channel. To support the 'RGB' color scheme, all encoded values were normalized from 0 to 255.

Using the encoded images, we trained our image-based deep learning model using a GAN-based architecture. GANs have been recognized as the best performing type of generative network[19]; a class of neural networks that can directly find multiple solutions to a given problem. Other types of networks that fall in this category include variational autoencoders (VAEs)[52] and mixture density networks (MDNs)[53]. Recent developments in GAN technology have led to numerous GAN-variants, including but not limited to: the Self-Attention GAN (SAGAN)[36], Deep Regret Analytic GAN (DRAGAN)[37], StyleGAN[38], Wasserstein GAN (WGAN)[39], and the Least Squares GAN (LSGAN)[40]. Here, as an initial proof of concept, we tested our framework using a modified cDCGAN architecture, as shown in Figure 2a. cDCGANs have previously been used to generate domain-specific images in response to input conditions.[41-43] Implemented in the PyTorch framework, the cDCGAN consists of a generator and a discriminator. Initially, batches of absorption spectra ($y$) are fed into the generator, along with a latent vector ($z$), to generate 'fake' images ($G$) that are similar to the 'real' images ($x$) from the training set. The latent vector is sampled from a random uniform distribution and allows the generator to map a probability distribution to a design space, thereby enabling a one-to-many mapping.[26] Both *G* and *x* are then fed into the discriminator (*D*), which attempts to distinguish the generated images from the real. Thus, the generator is trained to produce convincing images that deceive the discriminator, while the discriminator is trained not to be deceived — a competition which leads to the joint and stepwise improvement of both networks via their loss functions. These loss functions are calculated using the binary cross-entropy criterion, and the complete model interaction is represented as:

$$min_G \; max_D \; l(G,D) \; = \; E_{x \rightarrow p_{data}(x)}\{log \; D(x,y)\} \; + \; E_{z \rightarrow p_z(z)}\{log(1 - D(G(z,y))\}, \quad (1)$$

where *E* is the expected result, $p_{data}(x)$ is the training data distribution, $p_z(z)$ is the latent vector distribution, $log(D(x,y)) + log(1-D(G(z,y))]$ is the discriminator loss ($L_D$), and $log(D(G(z,y)))$ is the generator loss ($L_G$). During training, the objective is to maximize $L_D$ and $L_G$. We note that our definition of the $L_G$ differs from the original GAN implementation, where $log(1-D(G(z,y))$ is minimized instead, since this was shown to not provide sufficient gradients.[53,54] To improve the performance of the cDCGAN, we applied one-sided label smoothing and mini-batch discrimination.[44,45] Unlike previous cDCGAN implementations, our approach relies on adversarial training without explicitly guiding the generator towards known images,[21] thereby achieving a greater degree of generalization that is unconstrained by pre-existing images. Over 40 different cDCGAN architectures were trained through extensive hyperparameter tuning, and the optimized architecture can be found on Figure S3. Several alternative parameter-encoding schemes were also trained and presented in Figure S4, where models trained on several neurons (to represent encoded properties) were compared to models trained using the entire 'RGB' channels. The validation losses of each method are reported in Table S1 and S2, and the color-encoding approach is shown to exhibit the best performance among the tested encoding schemes. After training the cDCGAN, we developed an image processing workflow to convert the generated images into full 3D metasurface designs (Figure 2b). In this workflow, the material property ($\omega_P$ or *n*) and thickness values (t) are calculated by taking the average pixel-values in their respective channels (based on structure classification), then reversing the normalization performed in the encoding step. Additional details regarding this process can be found in the Supporting Information.

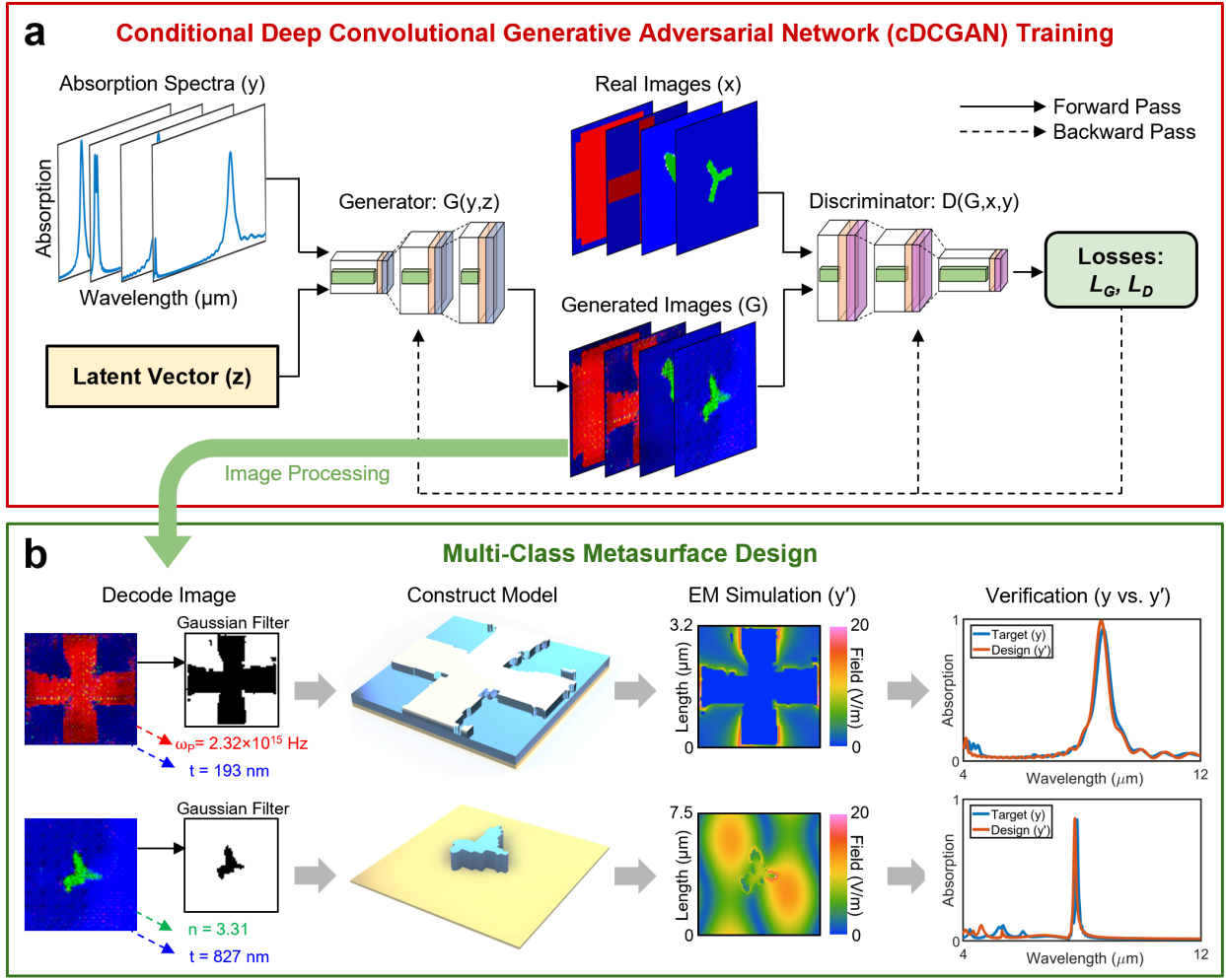

**Figure 2.** Schematic of the cDCGAN training and design process. (a) Both the generator and discriminator are neural networks that train in tandem to maximize the generator's accuracy. (b) After training, the generator can be used for multi-class metasurface inverse design. Images synthesized by the generator are decoded to construct 3D models of metasurfaces with unique material and structural parameters. The generated structures are then simulated to verify their adherence to the input target spectra.

In the GAN-metasurface design process, new materials were specified in the EM simulation software using the generated $\omega_P$ or $n$ values. We note that new materials created in this manner may not be compatible with fabrication schemes which rely on conventional materials. However, the presented material definition scheme allows the model to freely predict a continuum of material properties that are otherwise lost or disregarded due to categorical approximations, which enables a wider range of material property-driven designs. For example, metamaterials using dielectrics embedded with custom nanoparticle formulations can yield materials with effective refractive indices that can be deterministically tuned.[46-48,56] Prior studies have also employed nanoscale metallic alloying to achieve tailored plasma frequencies.[57] Highly granular material-level predictions, as we show are possible here, would therefore enable additional degrees of freedom for materials optimization, which may in turn yield novel optical responses.

      We evaluated the performance of our trained cDCGAN and image processing method by inputting a set of absorption spectra (coupled with randomly sampled latent vectors) and analyzing the resulting designs. Since the GAN may produce a distribution of designs with potentially varying degrees of accuracy,[8,51] ten different latent vectors were generated for each target spectrum, which were then used as inputs to the network. Each design is verified using numerical simulation, then the design (and corresponding latent vector) with the lowest mean-squared error to the target is reported as the final design. Figure S7 shows the distribution of designs (across different latent vectors) for several input targets, where we observe that each design variant has over 90% accuracy in comparison to the input target. Following this procedure, Figure 3 presents a series of tests performed with inputs that originate from the validation dataset (10% of the training dataset). Here, the blue lines represent randomly selected inputs (across both classes of structures), and the orange lines are the simulated spectra of the cDCGAN-generated designs. Images of the corresponding structures (direct outputs of the network) are shown to the right of each plot. Below each image are the associated material property ($\omega_P$ or $n$) and dielectric thickness values which are derived from the aforementioned decoding scheme. Figure S8 shows the equivalent results for inputs from the training dataset.

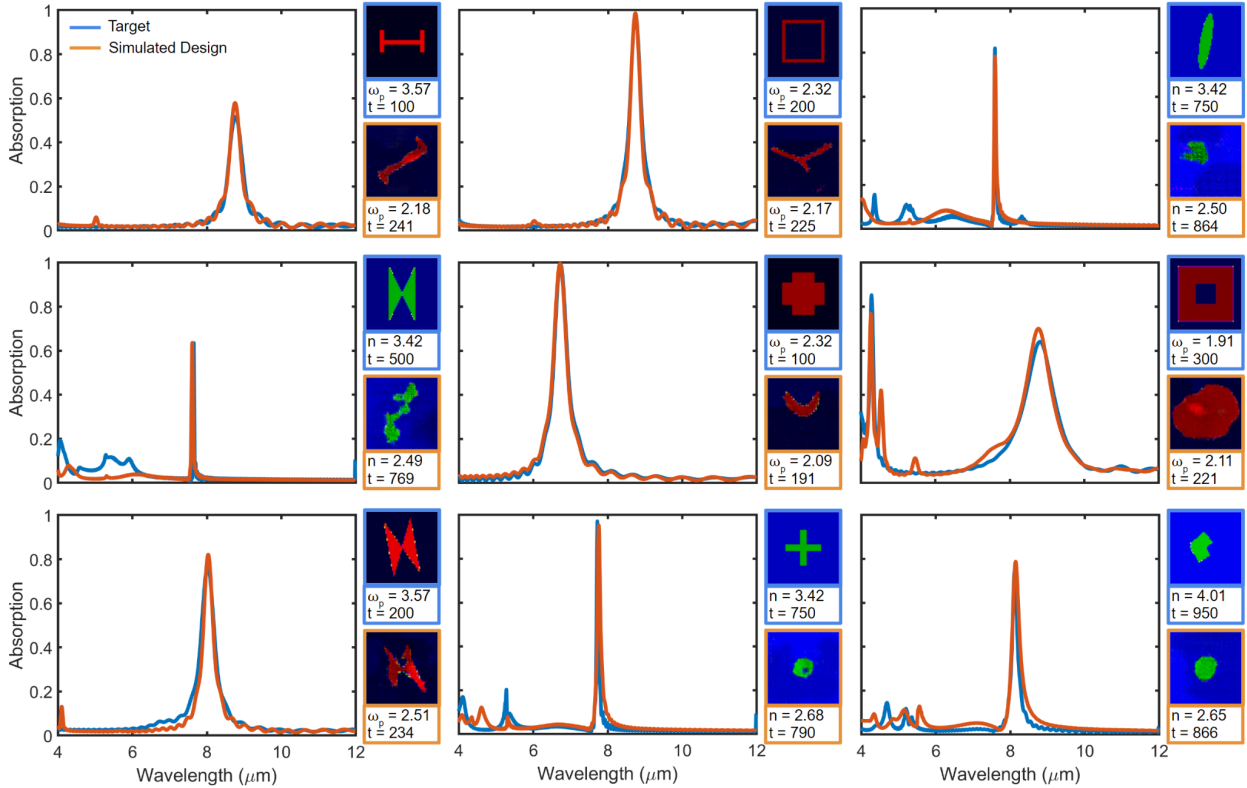

**Figure 3.** Randomly selected absorption spectra from the validation dataset (blue) which were designated as input targets for the cDCGAN. The simulated spectra of the cDCGAN-synthesized designs (orange) are plotted alongside the targets for comparison. Images representing the respective structures are shown to the right of each plot, with material and thickness information below each image. Units for plasma frequency ($\omega_P$) values are in PHz and thicknesses (t) are in nanometers. The results here reveal that the network can identify the underlying relationships between structure, material, metasurface class, and optical response to provide new yet accurate solutions that extend beyond the known designs.

We observe that in each test case, the network predicted the class of structure that corresponds with its particular type of spectral response. Specifically, when Fano-shaped spectra of various hybrid dielectric structures were passed into the network, the network exclusively generated hybrid dielectric structures (or green-blue images). Similarly, Lorentzian-shaped inputs yielded only MIM structures (or red-blue images). The generated images suggest that the network was capable of: 1) learning the distinguishing features and optical responses between the two explored classes of metasurfaces, and 2) using this information to predict the appropriate class based on the nature of the input spectra. In addition, across a wide range of input spectra, we observe that the network synthesized designs that are noticeably different from the known structures (either in resonator shape or property/thickness). Despite this difference, the generated designs exhibit responses that strongly match the input targets. Thus, these results show that our network is not simply mimicking designs from the training dataset. To a degree, the cDCGAN is capable of learning the underlying relationships between structure, material, metasurface class, and optical response to provide new yet accurate design solutions that extend beyond the training data.

To assess our network's ability to solve arbitrarily-defined design problems, we tested the network using 'hand drawn' target spectra. These targets are derived from the Fano resonance and Lorentzian distribution functions and have no associated design or structure. We evaluated the cDCGAN's performance across a wide range of inputs by using each function to create 200 spectra with amplitudes ranging from 0.5-0.9, and resonance wavelengths ranging from 5-9 $\mu$m, for 400 total test spectra. Figure 4a and Figure 4b show several results of the Fano-shaped and Lorentzian-shaped targets, respectively, where a strong match between the targets and simulated designs can be observed. A statistical evaluation of the entire test dataset is reported in Figure 4c (for the Fano-shaped targets) and Figure 4d (for the Lorentzian-shaped targets). Here, the histograms illustrate the number of test spectra which reside in specific MSE value ranges. Dashed-red lines indicate the average mean-squared error (MSE) of the Fano-shaped and Lorentzian-shaped targets, which equal to approximately $8.5 \times 10^{-3}$ and $2.9 \times 10^{-3}$, respectively. Through these plots, we note that the accuracy of the Fano targets is lower than the accuracy of the Lorentzian targets. However, further analysis of the training dataset (Figure S2) and the individual test results (Figure S5) reveal that the low-accuracy regions of the Fano-shaped structures correspond to regions that are not well-represented by the training data, whereas the high accuracy of the Lorentzian-shaped spectra can be explained by the wide spectral range of the MIM structures. Therefore, the performance of the Fano-shaped designs can potentially be improved by expanding the training data and design space.

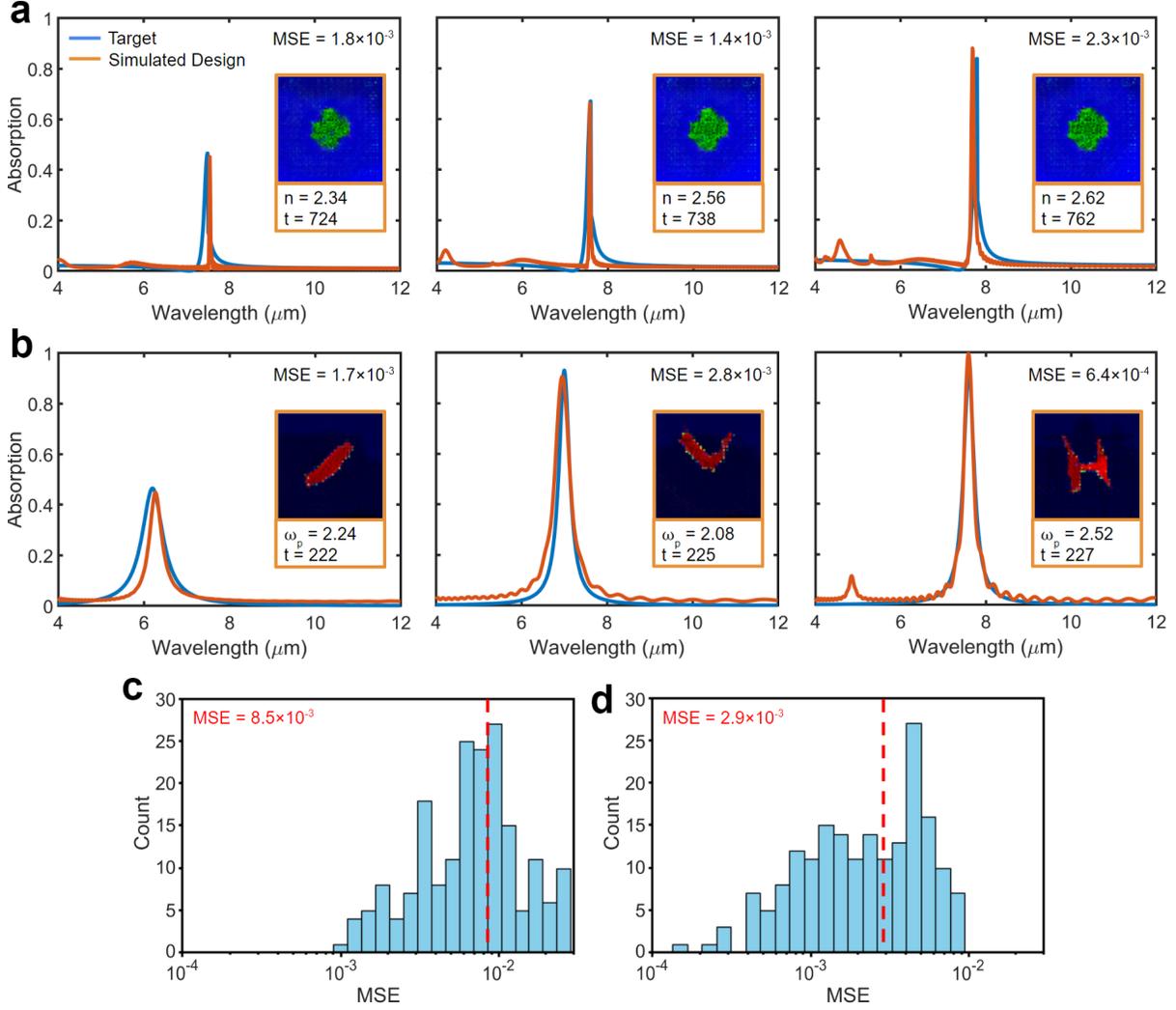

**Figure 4.** cDCGAN response to arbitrary 'hand drawn' targets for which there are no corresponding structures. The inset images show the synthesized images with material and structural information. (a) For Fano-shaped and (b) Lorentzian-shaped input targets, various hybrid dielectric and MIM structures with matching simulated responses are produced, respectively. Units for plasma frequency ($\omega_P$) values are in PHz and thicknesses (t) are in nanometers. Statistical analyses across the entire test dataset (400 total spectra) for the (c) Fano-shaped and (d) Lorentzian-shaped targets.

In principle, the 'one-to-many' mapping capabilities of GANs allow the deep learning model to generate multiple answers to a given problem. In the context of photonics design, this 'one-to-many' feature could provide an assortment of design options from which the designer can select from. Accordingly, to harness the full potential of our property-embedded cDCGAN, we evaluate and report the network's ability to generate multiple designs for a single target spectrum.

To ensure consistency, this 'diversity test' was performed on several target spectra. As seen in Figure 5a and Figure 5b, we queried the cDCGAN with Fano-shaped and Lorentzian-shaped spectra, respectively. For each spectrum (shown in their individual plots), a second query was performed after resampling the latent vector and slightly perturbing the starting spectrum. While not perturbing the spectrum still produced unique results on the second run (as shown in Figure S7), adding small perturbations (less than 0.01 shifts in amplitude at various wavelengths) increased the overall uniqueness of the new designs. It can be observed that for each of the Fano-shaped and Lorentzian-shaped inputs, the network is able to generate two designs with distinct resonator geometry, material properties, and/or dielectric thicknesses. Importantly, though the designs have varying levels of differences, their absorption spectra remain approximately the same. The diversity of 'one-to-many' structures for a target spectrum is tied to the available shapes and materials that the network was able to learn from, and allows us to make use of the non-uniqueness problem that is traditionally a limiting factor in inverse design approaches in photonics. A training dataset with a larger variety of materials and geometries could certainly yield a wider panel of designs for a given target, thereby providing end-users a range of materials and geometric designs that can deliver the same spectral response.

While the presented inverse design framework was intended to generate arbitrary material predictions as a means to enable additional degrees of freedom for geometry and materials optimization, a key limitation of the presented approach thus far is that constituent materials with arbitrarily-defined properties are generally more difficult to fabricate or synthesize than conventional materials. Accordingly, to enhance the capabilities of the proposed framework in terms of their fabricability and accessibility, we demonstrate that the GAN can be used with a look-up table to substitute the predicted material properties with the closest properties derived from standard materials (shown in Figure 6). In particular, Figure 6a shows a series of tests where the input targets are Fano-shaped spectra. Here, the GAN predicted arbitrary geometries, thicknesses, and refractive index values of 2.48, 2.32, and 2.58 (from left to right). We observe that the simulated structures match well with the target responses (as previously demonstrated).

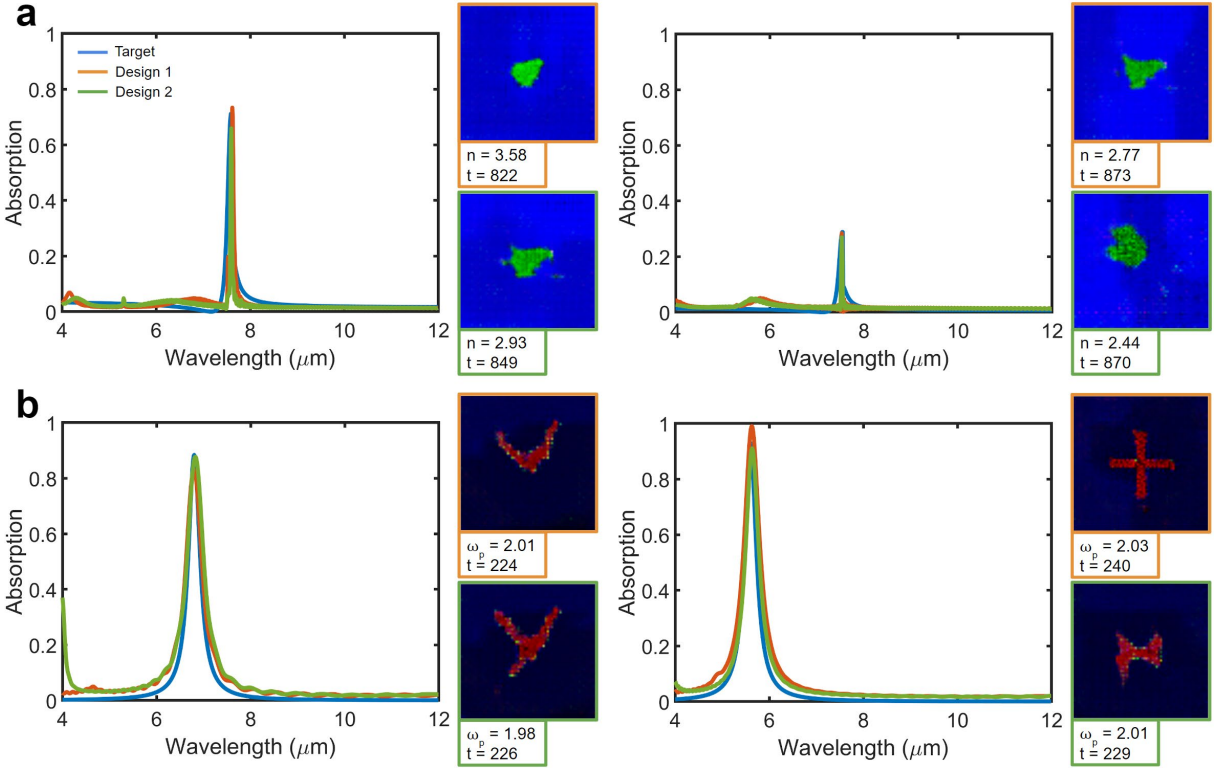

**Figure 5.** Demonstration of the 'one-to-many' mapping capabilities of the cDCGAN. Multiple structures with different materials and designs can be generated for a given (a) Fano-shaped or (b) Lorentzian-shaped target spectrum. Units for plasma frequency ($\omega_P$) values are in PHz and thicknesses (t) are in nanometers.

Next, to implement the look-up table, we substitute the GAN-generated values of *n* with those of the closest materials found in a publicly-available database,[61] including: CdSe (*n*=2.44), GaSe (*n*=2.38), and CdTe (*n*=2.68).[58-60] In Figure 6b, we perform a similar set of tests with Lorentzian-shaped spectra, where the predicted materials are substituted with Au and Ag.[31,32] In both cases, after repeating the simulations, we observe that the material approximations maintain ~90% accuracy in comparison to the GAN's true predictions. Thus, we demonstrate an alternative approach at using our inverse design framework to achieve designs with greater accessibility (while maintaining reasonable accuracy). We also note that some materials identified through this approach are unique and do not exist in the training dataset (CdSe, GaSe, and CdTe). However, by virtue of the GAN-based approach outputting a new material parameter (refractive index or plasma frequency) as its prediction, we are able to identify other materials (beyond the training data) that can meet the requirements of a newly sought target. We believe this highlights a notable strength of our approach, because class-based machine learning-based methods are restricted to predicting material categories that are only available in the training dataset. As we demonstrate here, our approach enables a new degree of generalization and design flexibility by allowing practitioners to access more materials than those represented by the training data. While the particular examples we presented show that the GAN predicts values which fall within the range of real materials, we acknowledge that the GAN may also predict properties beyond the current scope of conventional materials. However, we expect the accuracy of such material approximations to improve as material libraries, and material accessibility in general, continue to develop and grow.

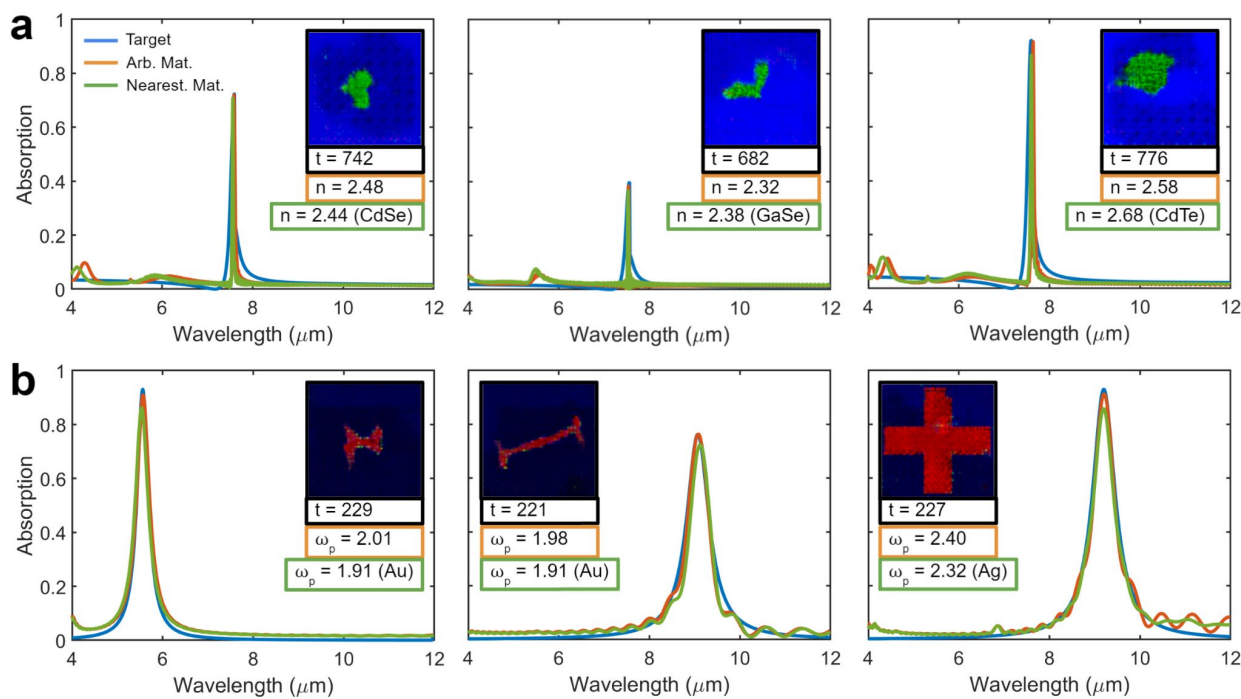

**Figure 6.** Applying similar materials to the cDCGAN predictions to increase fabricability. Comparison between (a) Fano-shaped and (b) Lorentzian-shaped input targets (blue). Units for plasma frequency ($\omega_P$) values are in PHz and thicknesses (t) are in nanometers. Simulated results reveal that material approximations (green) maintain ~90% accuracy in comparison to the GAN-predicted materials (orange).

## Conclusions

In summary, we present a deep learning-based photonics design framework that enables the simultaneous prediction of metasurface topology, material properties, and out-of-plane structural parameters across multiple classes of metasurfaces. Our framework is centered on a conditional deep convolutional generative adversarial network (cDCGAN) and a multiparametric-encoding strategy in which the colors of an image are encoded with various material and structural properties. By accounting for the global parameters which govern the optical behavior of metasurfaces (material, structure, and device class or fabrication process), our approach overcomes the key limitations of previously-demonstrated generative models, where only a few of the aforementioned design criteria were considered. Evaluation of our model's performance reveals that it is capable of generating not only accurate and distinct solutions from the training and validation datasets, but also multiple design alternatives and material recommendations for a single target by taking advantage of the 'one-to-many' mapping capabilities of GANs. To account for potential fabrication or material constraints, a property-based look-up mechanism can be paired with the model's predictions to identify readily-available materials that serve as reasonably-accurate substitutes. The presented encoding scheme is easily adaptable to existing generative models that are integrated with optimization algorithms.

Though only two classes of metasurfaces were explored in this study (metal-insulator-metal and hybrid dielectric resonators), we believe that the results here validate the feasibility of a deep learning-based global photonics design solution aimed at describing all physical aspects of a structure. Alternative encoding schemes with greater complexity, such as higher-dimensional tensors, may therefore be employed to capture more categories of photonic designs as well as more information regarding a structure's physical properties. To achieve a more generalized inverse design framework, future studies may directly incorporate other fundamental optical properties of materials (e.g. real and imaginary refractive indices, magnetic permeability, etc.) into the model. In this regard, a multi-pole Lorentz-Drude oscillator model with multiple parameters can also provide higher-accuracy fits over alternative wavelength ranges. More broadly, the presented methodology can be adapted to a wide range of materials design problems, including mechanical metamaterials and other synthesis-driven design challenges. Thus, our proposed framework offers a path towards a global machine learning platform that can allow practitioners and researchers to identify optimal combinations of materials, geometric parameters as well as device categories to meet complex and demanding performance goals in a range of physical systems.

# Acknowledgements

This work was supported by the Sloan Research Fellowship from the Alfred P. Sloan Foundation.

# Conflicts of interest

The authors declare no conflicts of interest.

# Supporting Information

**Training Dataset**

The training dataset for deep learning consists of precisely 18,770 metasurface unit cell designs (12,632 MIM and 6,138 hybrid dielectric structures). These designs were derived from seven starting shape templates: cross, square, ellipse, bow-tie, H, V, and tripole-shaped. As shown in Figure S1a, parameter sweeps were performed on each shape (for the MIM structures) to produce geometric variations. The tabulated parameter sweeps are captured within 3.2×3.2 $\mu m^2$ unit cells. For the hybrid dielectric structures, the same parameter sweeps were scaled by ×2.34 $\mu m$, and the unit cell dimensions for this group of structures were 7.5×7.5 $\mu m^2$. Both sets of structures are represented as 64×64×3 pixel images. Figure S1b shows several image pairs, which illustrate examples of finalized metasurface designs from each shape template. Images on the left are the color-encoded images used for deep learning, and images on the right represent the corresponding 3D physical models. As described in the main text, the colors on the image are used to indicate the metasurface class, resonator geometry, material choice, and dielectric thickness. Specifically, each structure possesses a 100 nm thickness gold substrate. For each MIM structure, the metal resonator is a 100 nm layer of gold, silver, or aluminum, while the dielectric material is $Al_2O_3$ with a thickness of 100 nm, 200 nm, or 300 nm. For each hybrid dielectric structure, the dielectric resonator is zinc selenide, silicon, or germanium, and its thickness is 500 nm, 750 nm, or 950 nm.

Full-wave simulations were performed on each structure (under p-polarization at normal incidence) to produce a corresponding 800-point absorption spectrum across the mid-infrared wavelength range. Low quality designs which exhibited 'flat' (maximum absorption is less than 0.2) or 'noisy' (mean-squared error, or MSE, between the spectra and its average is greater than 0.05) spectral responses were removed from the training set to maximize the model's utility and performance. Figure S2 illustrates the peak absorptions and resonance wavelengths of all the spectra training dataset, organized by full width at half maximum (FWHM). The distribution of absorption spectra reveals that the MIM structures (FWHM >= 0.4 $\mu m$) cover a wide range of peak amplitudes and resonance wavelengths from 4-10 $\mu m$, while a majority of the hybrid dielectric structures (FWHM <= 0.2 $\mu m$) exhibit responses from 7.5-8.5 $\mu m$. The range of responses in the training dataset may be extended in future studies to enhance the network's predictive capabilities. On a distributed high-performance computing cluster with four dedicated compute nodes per simulation, where a node has a minimum of four 64-bit Intel Xeon or AMD Opteron CPU cores and 8 GB memory, each FDTD simulation took approximately 5 minutes to complete. Therefore, our training dataset equates to approximately 65 days of simulation time.

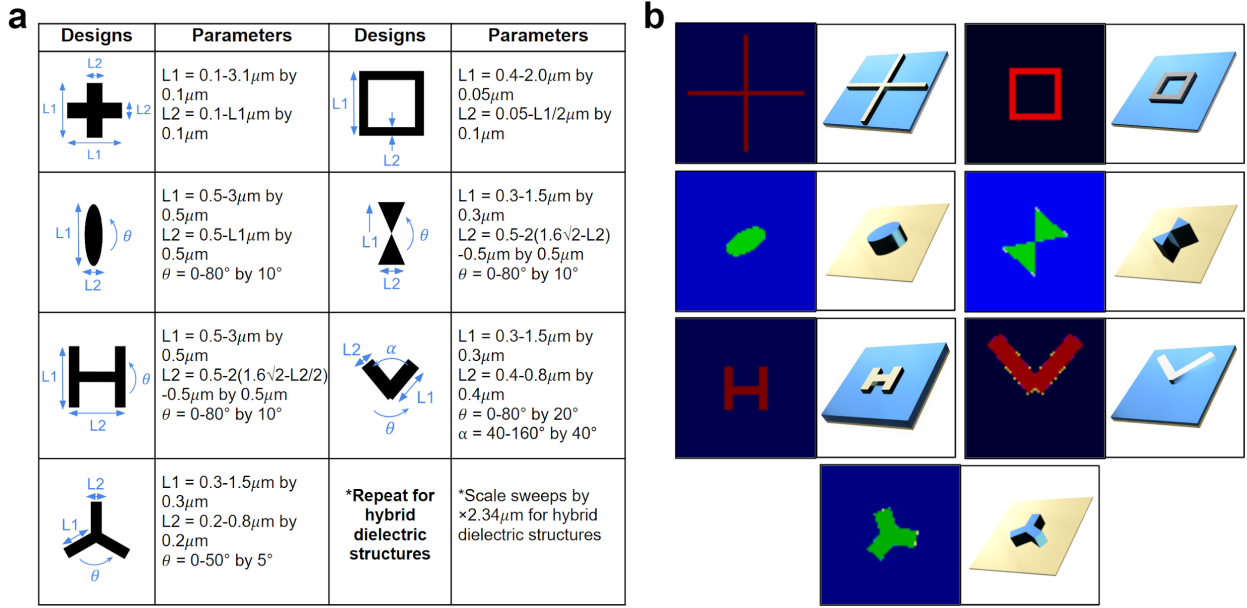

**Figure S1.** (a) 2D images of template shapes used to derive unit cell designs for both classes of metasurfaces. The range of variation allowed for each parameter is listed to the right of the associated shape. (b) Color-encoded 2D images (left) representing metasurface resonators. Images with red colors represent MIM structures while images with green colors represent hybrid dielectric structures. Shades of blue represent the dielectric thickness for both classes of structures. 3D models of each resonator are shown to the right of their 2D representation.

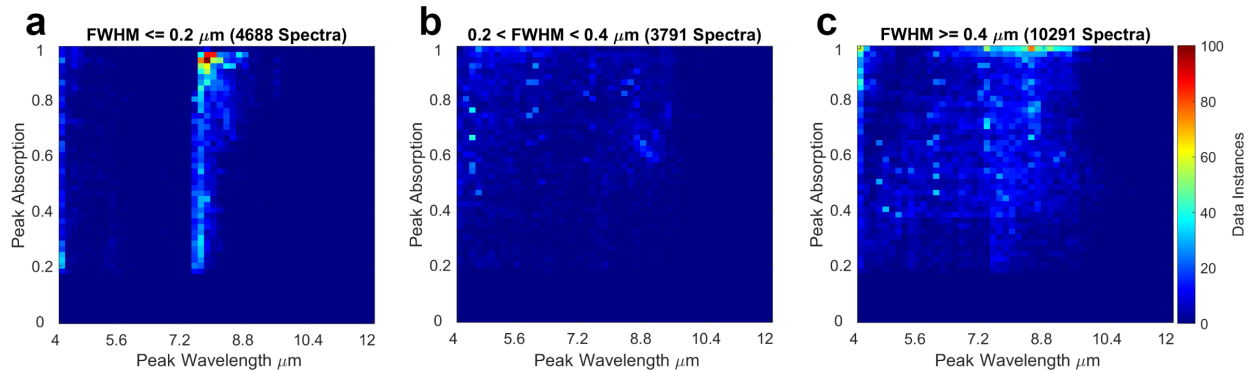

**Figure S2.** Visualization of peak absorption and wavelength values in the training set as distribution heatmaps. Regions of higher red-intensity indicate increased data instances within this range. The individual spectra within the training set are sorted into three subsets based on the FWHM of their absorption peak: (a) FWHM less than or equal to 0.2 μm (b) between 0.2 μm and 0.4 μm, and (c) greater than 0.4 μm.

**Network Architecture Design and Optimization**

Implemented in the PyTorch framework, the cDCGAN consists of two networks: a generator and a discriminator (illustrated in Figure S3). The optimized generator contains five transposed convolutional layers (with 1200, 1024, 512, 256, and 128 input channels or feature maps), while the discriminator has five convolutional layers (with 6, 64, 128, 256, and 512 input channels or feature maps). Each transposed convolutional layer in the generator is followed by a batch normalization and ReLU (rectified linear unit) activation layer, instead of the final layer, where a Tanh (hyperbolic tangent) activation is used. Similarly, in the discriminator, each convolutional layer is followed by a batch normalization and Leaky ReLU layer, and the final layer possesses a Sigmoid activation. At the generator input layer, the 800-point absorption spectra are concatenated with 400-point latent vectors to yield 1200-point input vectors. For the discriminator input, the absorption spectra are passed through a fully-connected layer and reshaped into a 64×64×3 matrix. These matrices were then concatenated with the real and generated images to form 64×64×6 inputs for the discriminator. Model training was performed on an NVIDIA Titan RTX GPU and took approximately 30 minutes to complete.

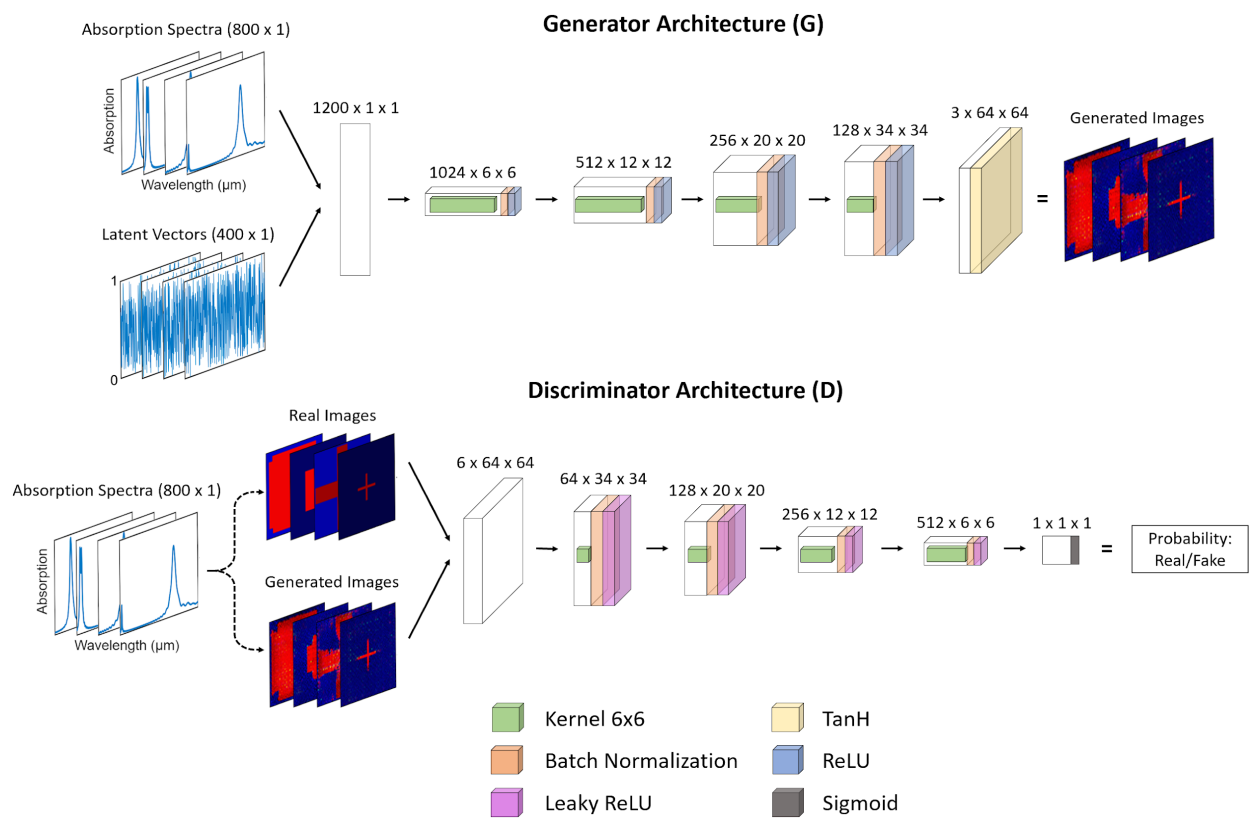

**Figure S3.** Schematic of the generator and discriminator architectures implemented in our cDCGAN model. Input and output types are shown for each layer along with the layer types and dimensions.

**Evaluation of Multiparametric Encoding Methods**

We tested the efficacy and performance of three different material and structural parameter encoding methods. Figure S4 shows the training progression of each encoding method at various epochs. As seen in Figure S4a, the first encoding method uses several neurons to represent the parameters by embedding them into a single row and column of pixels within the topological image (for each parameter). Specifically, in this encoding scheme, an initial 64×64 pixel image is converted to a 66×66 image, where the new rows and columns represent the material property and dielectric thickness of the metasurface design. The second method (Figure S4b), similar to the first, encodes the material and structure parameters as two rows and columns per parameter (yielding a 68×68 image). The third and final method that was investigated (Figure S4b) involves encoding the parameters into discrete 'RGB' channels of colored images, as described in the main text.

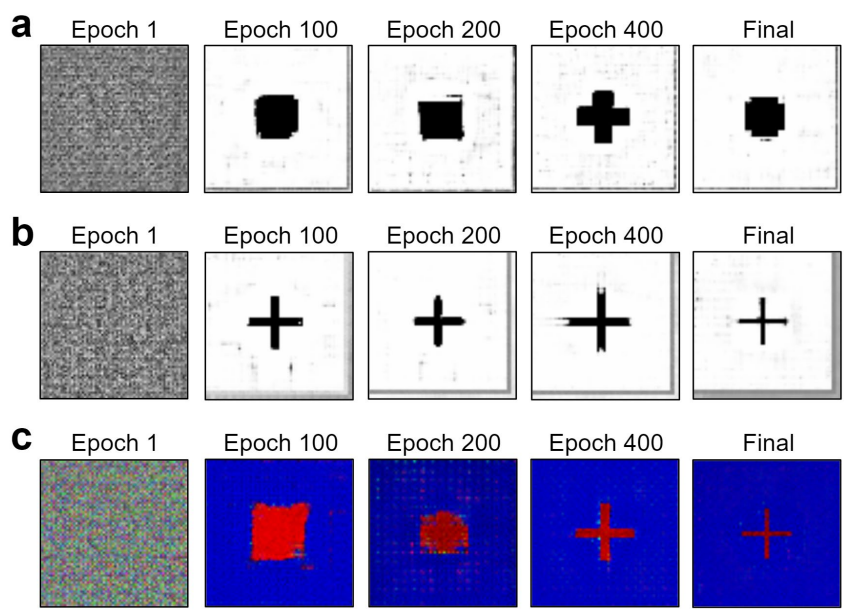

**Figure S4.** Examples of the generative model's training progress for three implementations of multiparametric encoding over hundreds of epochs. The tested implementations embedded parameter information as (a) a single-row and column vector concatenated with the 2D image, (b) double-row and column vectors concatenated with the image, and (c) normalized values within the 'RGB' channels of colored images.

Each encoding method was applied to the training dataset, and the corresponding datasets were separately used to train the cDCGAN. Hyperparameter tuning was performed via grid search, and the results for each dataset are presented in Table S1. Here, various feature maps, kernels, batch sizes, epochs, and miscellaneous pre- and post-processing steps were tested. Listed feature map sizes represent the lowest denomination of maps used in the intermediate layers (between the first and last layers) of the generator and discriminator. Reported losses are derived from the validation dataset.

**Table S1.** Hyperparameter optimization for various multiparametric encoding methods. Highlighted cells indicate the lowest validation loss for the corresponding method.

| Model | Feature Maps (G/D) | Kernel Size | Batch Size | Misc. | Epochs | Validation Loss (MSE) |
|---|---|---|---|---|---|---|
| colspan="7" | **Single Row/Column Encoding (MIM Only)** |||||||
| 1 | 32/16 | 4 | 128 | | 750 | 0.0264 |
| 2 | 64/32 | 4 | 128 | | 750 | 0.0215 |
| 3 | 64/32 | 4 | 128 | | 1000 | 0.025 |
| 4 | 66/66 | 4 | 16 | Gaussian Filter ($\sigma$=0.75) | 500 | 0.0269 |
| 5 | 66/66 | 4 | 16 | Gaussian Filter ($\sigma$=0.75) | 750 | 0.0332 |
| 6 | 66/66 | 4 | 128 | Gaussian Filter ($\sigma$=0.75) | 750 | **0.0127** |
| 7 | 66/66 | 4 | 128 | | 750 | 0.0135 |
| 8 | 66/66 | 6 | 16 | Gaussian Filter ($\sigma$=0.75) | 500 | 0.0351 |
| 9 | 66/66 | 6 | 16 | Gaussian Filter ($\sigma$=0.75) | 750 | 0.0405 |
| colspan="7" | **Double Row/Column Encoding (MIM Only)** |||||||
| 10 | 68/68 | 4 | 128 | | 750 | 0.0413 |
| 11 | 68/68 | 5 | 16 | | 250 | 0.0379 |
| 12 | 68/68 | 5 | 16 | | 500 | 0.0292 |
| 13 | 68/68 | 5 | 32 | | 500 | 0.0291 |
| 14 | 68/68 | 5 | 64 | | 500 | 0.0157 |
| 15 | 68/68 | 6 | 16 | | 500 | 0.0128 |
| 16 | 68/68 | 6 | 16 | Gaussian Filter ($\sigma$=1) | 500 | 0.0168 |
| 17 | 68/68 | 6 | 16 | Gaussian Filter ($\sigma$=0.75) | 500 | **0.0112** |
| 18 | 68/68 | 6 | 32 | | 500 | 0.0177 |
| 19 | 68/68 | 6 | 32 | Gaussian Filter (1) | 500 | 0.0183 |
| 20 | 68/68 | 6 | 32 | Normal Distribution (z) | 500 | 0.0281 |
| 21 | 68/68 | 6 | 32 | Noise 350 pts (z) | 500 | 0.0328 |
| 22 | 68/68 | 6 | 32 | Noise 450 pts (z) | 500 | 0.0294 |
| 23 | 68/68 | 6 | 68 | | 500 | 0.0247 |
| 24 | 68/68 | 6 | 128 | | 500 | 0.0211 |
| colspan="7" | **Color-Encoding (MIM Only)** |||||||
| 25 | 64/32 | 6 | 16 | Boundary Thresh. (0.2) + GF | 750 | **0.0044** |
| 26 | 64/32 | 6 | 16 | Boundary Thresh. (0.1) + GF | 750 | 0.0153 |
| 27 | 64/32 | 6 | 32 | Boundary Thresh. (0.2) + GF | 500 | 0.013 |
| 28 | 64/32 | 6 | 32 | Boundary Thresh. (0.1) + GF | 500 | 0.0142 |
| 29 | 64/32 | 6 | 32 | Boundary Thresh. (0.2) + GF | 750 | 0.0111 |
| 30 | 64/32 | 6 | 32 | Boundary Thresh. (0.1) + GF | 750 | 0.0116 |
| 31 | 64/32 | 6 | 64 | Boundary Thresh. (0.2) + GF | 500 | 0.0179 |
| 32 | 64/32 | 6 | 64 | Boundary Thresh. (0.2) + GF | 750 | 0.0285 |

We note that the highest-performing hyperparameters for a specific encoding method was not typically the optimal model for other encoding methods. As a result, each encoding method was optimized independently of prior models. To expedite our training efforts, only the MIM structures were used for the first round of optimization. Across all the explored encoding methods, we observe that the color-encoding approach yielded the lowest validation loss (0.0044) and highest performance. In addition to hyperparameter tuning, a Gaussian filter (GF) with binary thresholding offered substantial performance gains (where $\sigma$ is the standard deviation of the Gaussian kernel) and significant reduction in noise-related artifacts, while modifying the latent vector (*z*) size and distribution (uniform to normal) resulted in no noticeable improvements.

When training the cDCGAN with the color-encoded images, the discriminator frequently overpowered the generator and resulted in mode collapse. Thus, we reduced the size of the discriminator's layers in comparison to the generator to balance the two networks. Furthermore, we developed an image processing workflow to convert the generated images into full 3D metasurface designs. Here, each generated image is decoded into three components: a resonator-only image, a material property value, and a dielectric thickness value. The resonator image specifies the existence (black pixels) or absence (white pixels) of planar features. These pixels are obtained by determining the boundaries between major color gradients on the GAN-generated color images (*e.g.*, red-blue or green-blue transition points), thus a boundary conversion threshold was applied in order to find the exact transition points. Here, we determined that the optimum threshold was a fifth of the maximum resonator color intensity (shown as 0.2 in Table S1 and S2) in the red or green color channel. If a pixel position possessed a red or green pixel value that exceeded the threshold, then the existence of a physical structure was indicated here. Prior to determining these feature boundaries, a binary classification is performed by calculating the dominant class-specific color, which is used to classify the structure (if red pixels are greater than green, then the structure is MIM, and vice versa for hybrid dielectric). The purpose of this procedure is to filter any stray red pixels that may be intermingled with green and vice versa, and to assign the appropriate boundary conditions and unit cell dimensions to the FDTD model. As described in the main text, material property and thickness values are then calculated by taking the average pixel-values in their respective channels (based on structure classification), then reversing the normalization performed in the encoding step.

**Table S2.** Final hyperparameter optimization with the entire training dataset. The highlighted cell indicates the model with the lowest validation loss.

| Model | Feature Maps (G/D) | Kernel Size | Batch Size | Misc. | Epochs | Validation Loss (MSE) |
|---|---|---|---|---|---|---|
| **Color-Encoding (MIM + DM Only)** ||||||||
| 33 | 64/32 | 6 | 16 | Boundary Thresh. (0.2) + GF | 750 | 0.0128 |
| 34 | 64/32 | 6 | 16 | Boundary Thresh. (0.2) + GF | 1000 | 0.0094 |
| 35 | 64/32 | 6 | 32 | Boundary Thresh. (0.2) + GF | 500 | 0.0136 |
| 36 | 64/32 | 6 | 32 | Boundary Thresh. (0.2) + GF | 750 | 0.0086 |
| 37 | 64/32 | 6 | 32 | Boundary Thresh. (0.2) + GF | 1000 | 0.0135 |
| 38 | 64/32 | 6 | 32 | Boundary Thresh. (0.2) + GF | 700 | 0.0115 |
| 39 | 64/32 | 6 | 32 | Boundary Thresh. (0.2) + GF | 800 | 0.0125 |
| 40 | 128/64 | 6 | 16 | Boundary Thresh. (0.2) + GF | 500 | ==0.0076== |
| 41 | 128/64 | 6 | 16 | Boundary Thresh. (0.2) + GF | 750 | 0.0106 |
| 42 | 128/64 | 6 | 16 | Boundary Thresh. (0.2) + GF | 1000 | 0.0125 |
| 43 | 128/64 | 6 | 32 | Boundary Thresh. (0.2) + GF | 1000 | 0.0144 |

After identifying that the color-encoding strategy resulted in the highest design accuracy (or lowest validation loss), a second round of optimization was conducted on the entire training dataset. As shown in Table S2, with the optimized post-processing procedure determined in the previous section, the highest-performance model was trained with a kernel size of 6, batch size of 16, 128 base generator feature maps, 64 base discriminator feature maps, and for 500 epochs.

**Batch Testing**

We evaluated the cDCGAN's performance across a wide range of new inputs by creating 200 Fano-shaped and Lorentzian-shaped spectra with amplitudes ranging from 0.5-0.9, and resonance wavelengths ranging from 5-9 $\mu$m, for 400 total test spectra. Figure S5 illustrates a comparison of 50 individual responses within this 'batch' test. Each tiled plot is presented with 4-12 $\mu$m wavelength and 0-1 absorption axes limits. Here, we observe that the Fano-shaped responses (Figure S5a) are most accurate between resonance wavelengths of 7.5-8.5 $\mu$m, while the Lorentzian-shaped responses (Figure S5b) maintain strong matches across the entire test data range. Regions of low accuracy correspond to the areas that are not well-represented by the training dataset (shown in Figure S2). Therefore, the performance of the cDCGAN may be improved by expanding the training data and design space.

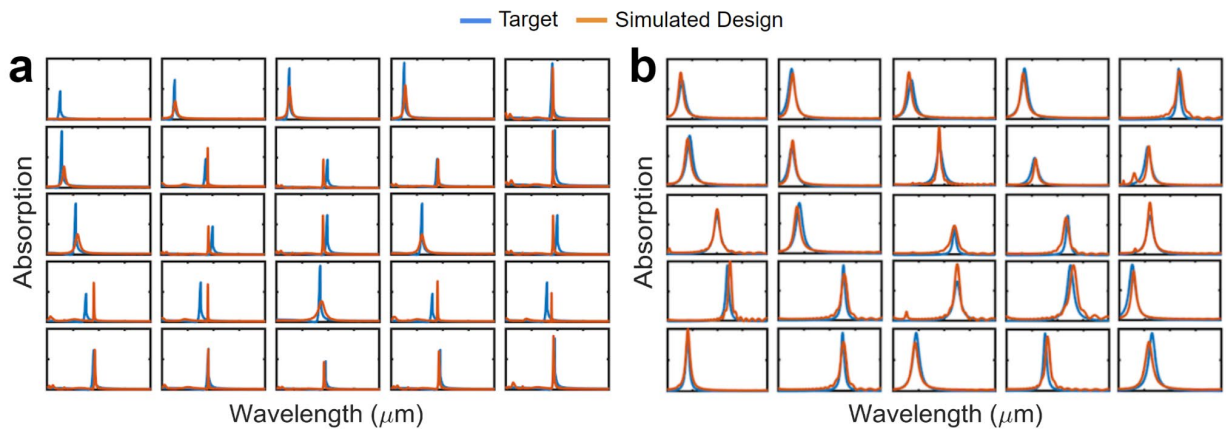

**Figure S5.** Test results of the cDCGAN using a diverse range of newly-constructed (a) Fano-shaped and (b) Lorentzian-shaped inputs. Blue lines represent input targets while the orange lines represent simulated designs produced by the cDCGAN.

**Designing Complex Alloyed Structures**

Though we limited this study to the application of uniform materials, we note that the devised color-encoding strategy is capable of representing spatially-varying material properties along the entire physical structure. This in turn sets the stage for future studies with much greater design complexity, such as 3D or complex metal alloy-based structures, with potentially greater control over the electromagnetic spectrum. A demonstration of this capability is shown in Figure

S6, where the different shades of color on the cDCGAN-generated MIM structures are converted into different metals (shown in the inset images) based on their individual plasma frequencies, rather than the average over the channel. Notably, the simulated alloyed structures yield similar responses to the uniform material structures, beyond which there are no distinguishable advantages in the particular design space that was explored. Therefore, future studies utilizing a wider range of dissimilar materials (as well as the application of fabrication constraints tailored towards alloy-based design) may produce device properties that extend beyond the uniform material domain to gradient-index materials.

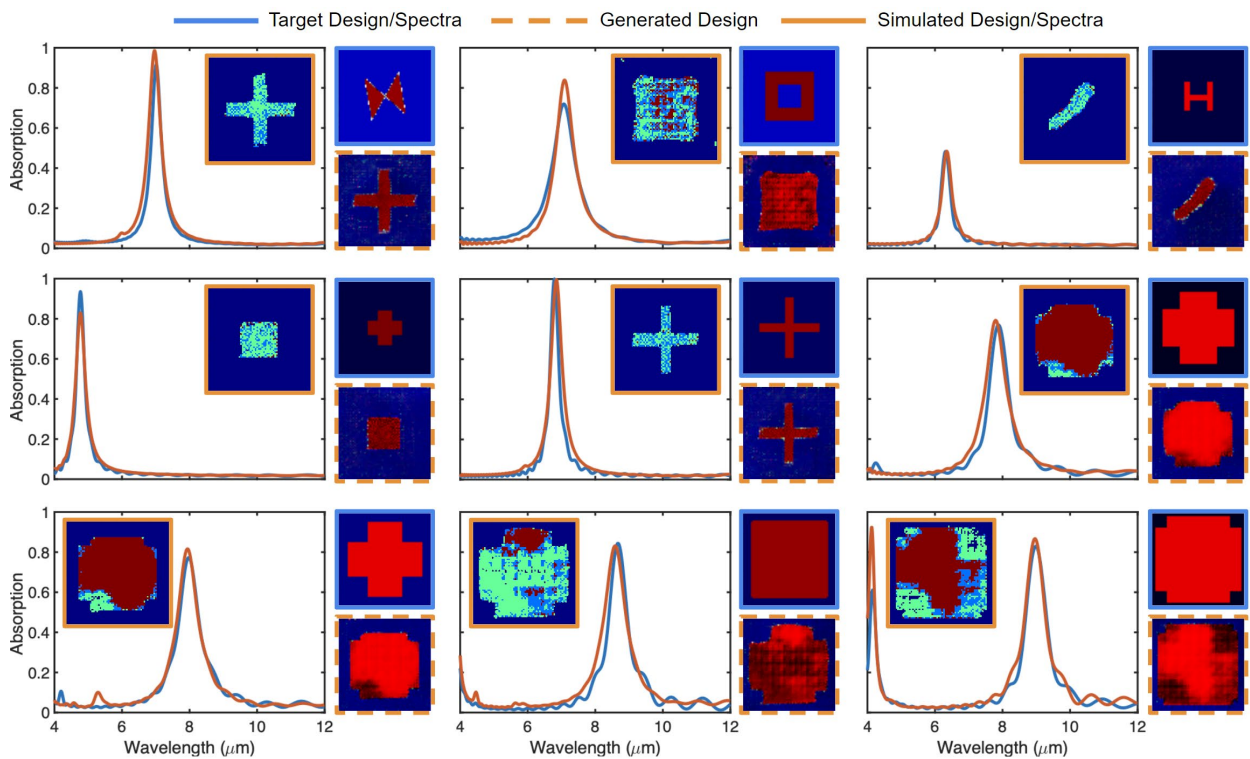

**Figure S6.** Demonstration of metal alloy-based structure design using the color-encoded cDCGAN. Blue lines represent the input spectra and reference designs. Dashed-orange lines represent the cDCGAN output, and solid-orange lines are alloyed structures created using the color gradients. We note that the simulated results match well with the input targets. The fabricability of these structures could potentially be improved with the addition of fabrication constraints such as minimum feature size.

**Latent Vector Sampling and Model Validation**

Since the GAN may produce a distribution of designs with potentially varying degrees of accuracy, ten different latent vectors were generated for each target spectrum, which were then used as inputs to the network. Each design is verified using numerical simulation, then the design (and corresponding latent vector) with the lowest mean-squared error to the target is used as the final design. Figure S7 shows the distribution of designs across different latent vectors for several input targets (a Lorentzian function centered at 7.2 μm and at 7.8 μm), where we observe that all the generated designs have over 90% accuracy in comparison to the input target. Following this procedure, Figure S8 presents a series of tests performed with inputs that originate from the training dataset. An equivalent analysis for the validation dataset can be found in the main text.

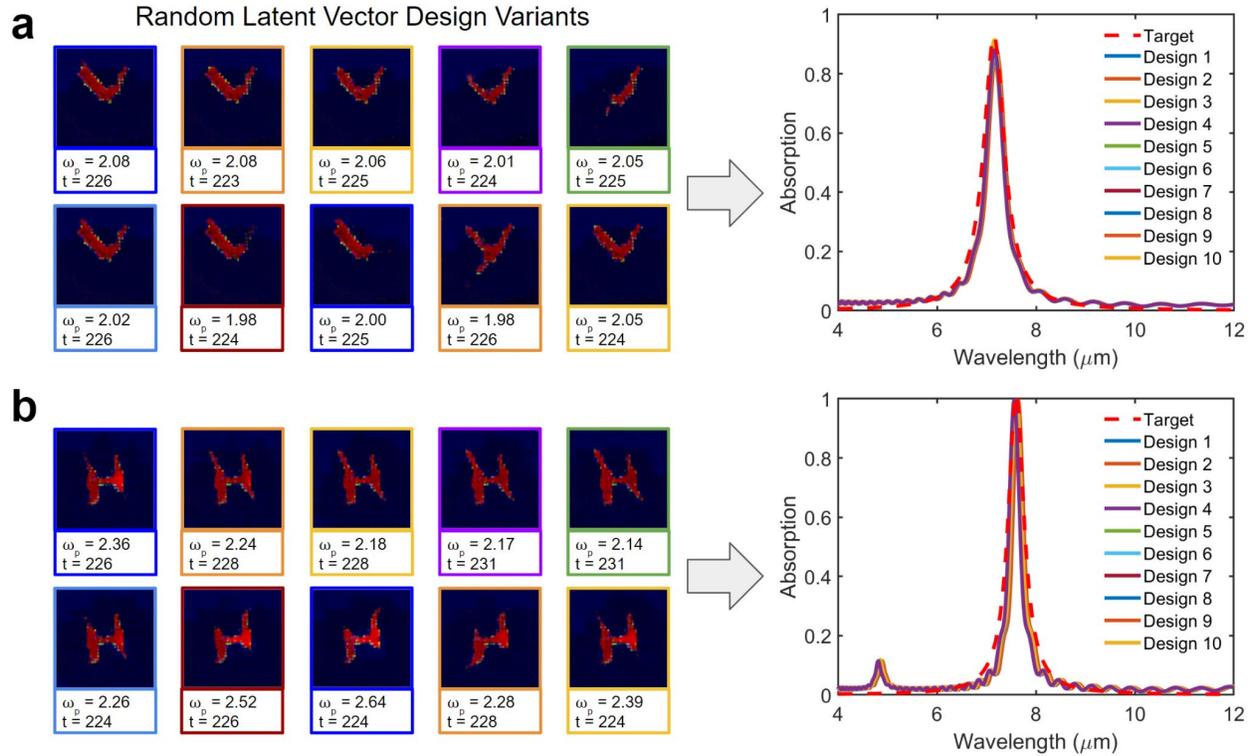

**Figure S7.** GAN-produced design variants achieved by pairing the target spectrum with 10 randomly-generated latent vectors. Input targets for Lorentzian functions centered at (a) 7.2 μm and (b) 7.8 μm are indicated by the dashed red lines. For each target, the design with the lowest mean-squared error is used as the final design.

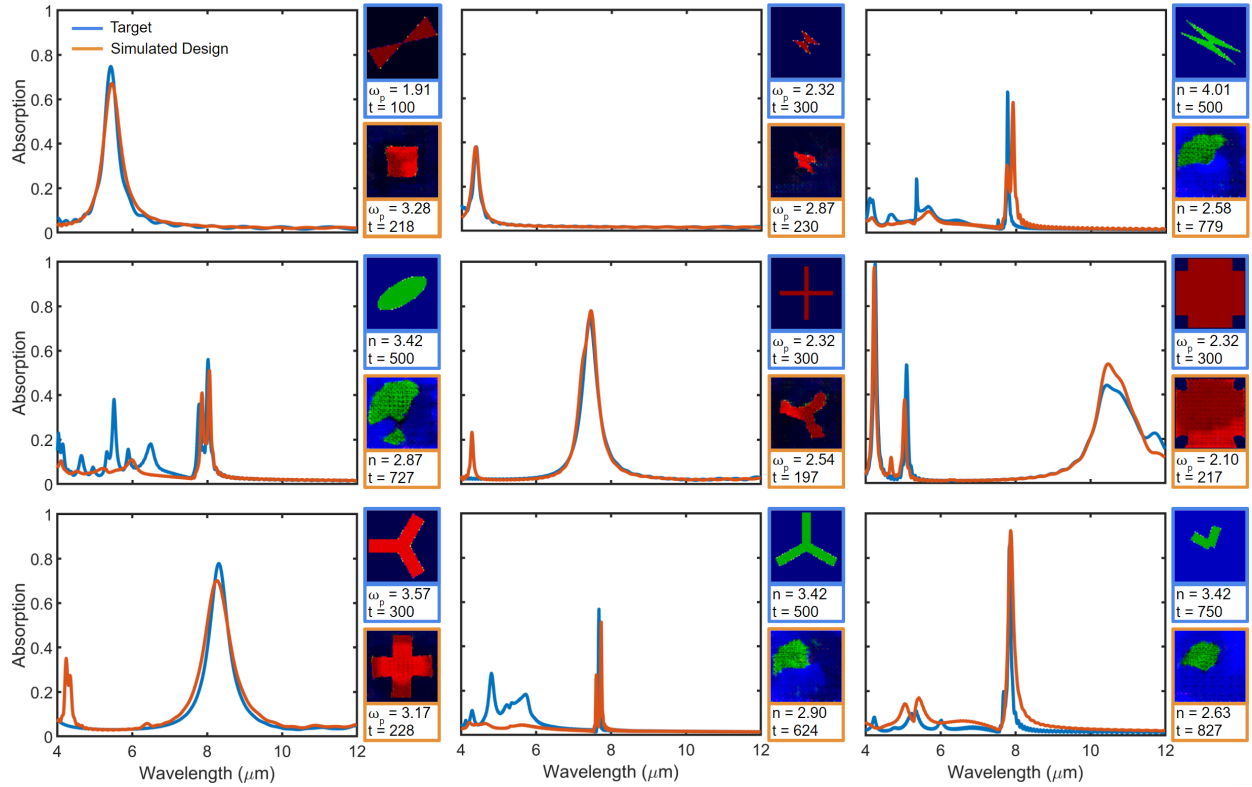

**Figure S8.** Randomly selected absorption spectra from the training dataset (in blue) which were designated as input targets for the cDCGAN. The simulated spectra of the cDCGAN-synthesized designs (in orange) are plotted alongside the targets for comparison. Images representing the respective structures are shown to the right of each plot, with material and thickness information below each image. Units for plasma frequency ($\omega_P$) values are in PHz and thicknesses (t) are in nanometers. The results here reveal that the network is not copying the training dataset, but to a degree, it is identifying the underlying relationships between structure, material, metasurface class, and optical response to provide new yet accurate solutions that extend beyond the training dataset.